\documentclass{iau} 
\usepackage{graphicx}

\title[TESS light curves of LMDEBs] %% give here short title %%
{TESS light curves of low-mass detached eclipsing binaries}

\author[K. G. He{\l}miniak et al.]   %% give here short author list %%
{Krzysztof G. He{\l}miniak$^1$,
 Andr\'es Jord\'{a}n$^{2,3}$, Nestor Espinoza$^{4,2}$ \\
\and Rafael Brahm$^2$}

\affiliation{
$^1$N. Copernicus Astronomical Center, Polish Academy of Sciences, 
\\ul. Rabia\'{n}ska 8, 87-100, Toru\'{n}, Poland \\ email: {\tt xysiek@ncac.torun.pl}\\
[\affilskip]
$^2$Instituto de Astrof\'{i}sica, Pontificia Universidad Cat\'olica de Chile \\
Av. Vicu\~{na} Mackenna 4860, 7820436 Macul, Santiago, Chile \\ 
[\affilskip]
$^3$Milleium Institute of Astrophysics\\
Av. Vicu\~{na} Mackenna 4860, 7820436 Macul, Santiago, Chile \\
[\affilskip]
$^4$Space Telescope Science Institute\\
3700 San Martin Dr., Baltimore, MD 21218, USA
}

\pubyear{2020}
\volume{354}  %% insert here IAU Symposium No.
\jname{Solar and Stellar Magnetic Fields: Origins and Manifestations}
\editors{A. Kosovichev, K. Strassmeier \& M. Jardine, eds.}
\begin{document}

\maketitle

\begin{abstract}
We present high-precision light curves of several M- and K-type, active
detached eclipsing binaries (DEBs), recorded with 2-minute cadence by the
{\it Transiting Exoplanet Survey Satellite} (TESS). Analysis of these curves,
combined with new and literature radial velocity (RV) data, allows to
vastly improve the accuracy and precision of stellar parameters with
respect to previous studies of these systems. Results for one previously
unpublished DEB are also presented.
%% add here a maximum of 10 keywords, to be taken form the file <Keywords.txt>
\end{abstract}

\firstsection % if your document starts with a section,
              % remove some space above using this command.
\section{Introduction}

Magnetic fields in low-mass ($<$0.8 M$_\odot$) stars affect the fundamental
stellar properties. In short-period, tidally locked binaries fast rotation of
components strengthens the magnetic field through a form of a dynamo
mechanism, enhances activity, and affects the observed radii and
effective temperatures, which has been observed in low-mass detached
eclipsing binaries (LMDEB) for decades. Several descriptions of this
phenomenon have been proposed, but we lack good quality observational
data and models of LMDEBs in order to validate or falsify them. In this paper
we present improved, precise results for four already studied and one unpublished 
LMDEB.

\section{Targets}

The presented sample consists of one system that was not studied to date, ASAS J125516-3156.7 (A-125), 
and four LMDEBs already described in the literature. These are:  
ASAS~J011328-3821.1 (\cite[A-011; He{\l}miniak \etal\ 2012]{hel12}), 
ASAS~J030807-2445.6 (\cite[AE~For; R\'o\.zyczka \etal\ 2013]{roz13}), 
ASAS~J032923-2406.1 (\cite[AK~For; He{\l}miniak \etal\ 2014]{hel14}), and
ASAS J093814-0104.4 (\cite[A-093; He{\l}miniak \etal\ 2011]{hel11}). 
All systems have component masses below 0.8~M$\odot$, and orbital periods shorter than 4 days. 
All are very active, with prominent spots, occasional flares, and H$\alpha$ emission lines. 
The four targets from literature usually have radii known with precision of $\sim$1.5-2\% at best. 
Masses of A-011 and A-093 were poorly constrained~($>$4\%).

\begin{table}
  \begin{center}
  \caption{Basic orbital and stellar parameters of the studied systems.}
  \label{tab}
 {%\scriptsize
  \begin{tabular}{l|ccccc}
\hline 
{\bf ASAS ID} & {011328-3832.1} & {030807-2445.6} & {032923-2406.1} & {093814-0104.4} & {125516-3156.7}\\
{\bf TIC}	& 183596242 & 88479623 & 144539611 & 14307980 & 103683084 \\
\hline
TESS Sector		& 2,3	&  4	&  4	&  8	& 10	\\
$P$ [d]			& 0.44559604(18)&0.918207(7)& 3.9809620(45)	& 0.897420(2)	& 3.0570393(44) \\
$K_1$ [km/s]	& 118.4(2.0)	& 118.3(5)	& 70.47(3)	& 127.55(68)& 73.34(5)	\\
$K_2$ [km/s]	& 162.9(3.3)	& 119.5(5)	& 77.16(5)	& 127.62(97)& 87.00(15)	\\
$i$ [$^\circ$]	&  87.5(1.5)	&  87.8(1)	& 87.37(3)	&  86.87(6)	& 87.56(7)	\\
$M_1$ [M$_\odot$]	& 0.597(28)	& 0.644(6)	& 0.696(1)	& 0.775(12)	& 0.7104(25)\\
$M_2$ [M$_\odot$]	& 0.434(17)	& 0.638(6)	& 0.6356(7)	& 0.774(10)	& 0.5989(13)\\
$R_1$ [R$_\odot$]	& 0.607(12)	& 0.674(7)	& 0.684(18)	& 0.774(6)	& 0.669(4)	\\
$R_2$ [R$_\odot$]	& 0.445(12)	& 0.617(10)	& 0.628(20)	& 0.771(6)	& 0.557(8)	\\
\hline
  \end{tabular}
  }
 \end{center}
\end{table}

\section{TESS photometry}
\begin{figure}
\begin{center}
\includegraphics[scale=0.43]{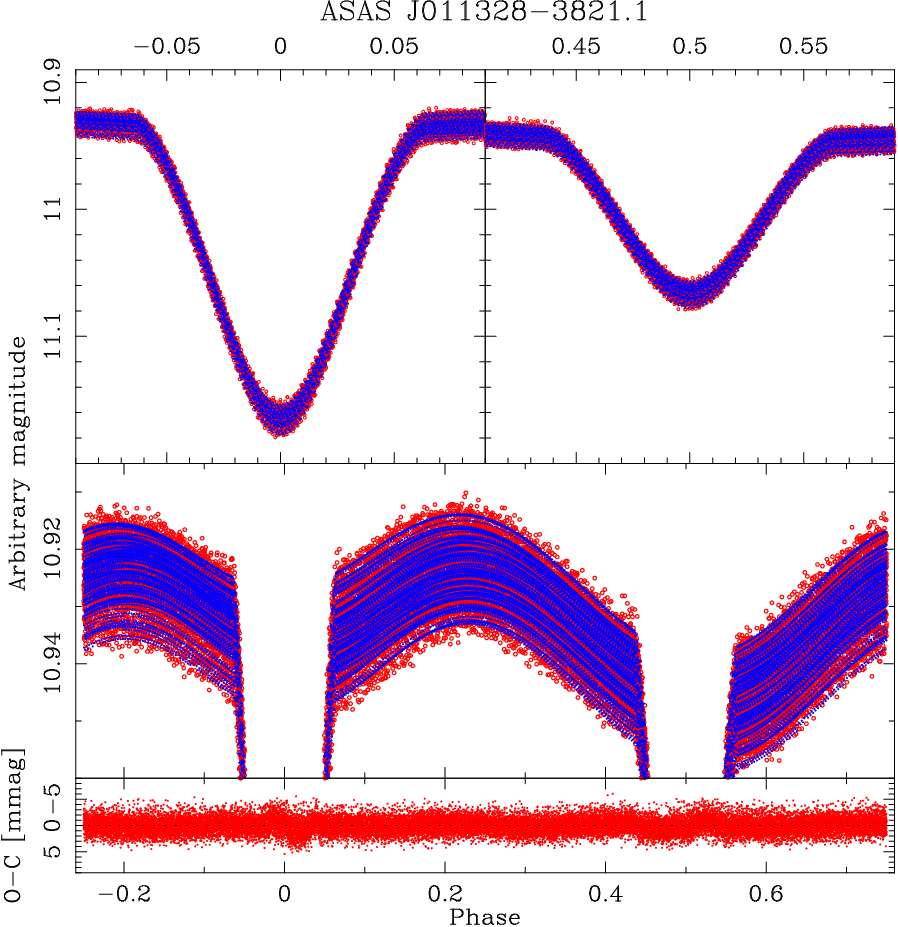}
\includegraphics[scale=0.43]{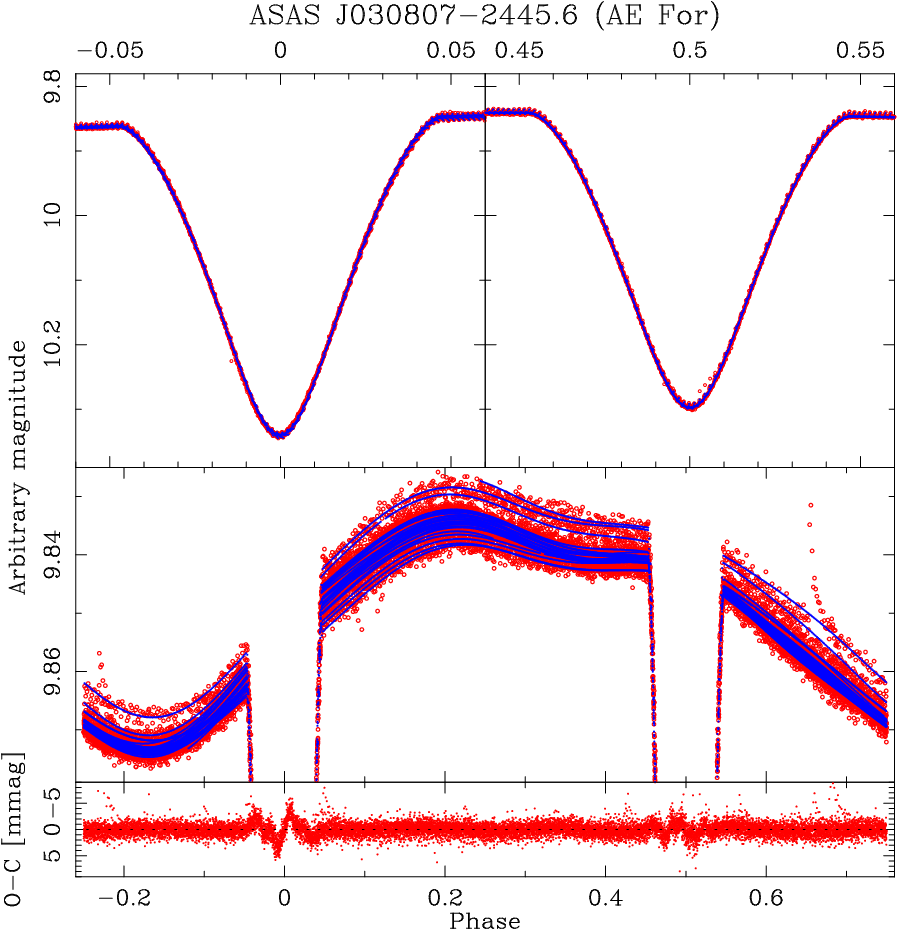}
\includegraphics[scale=0.43]{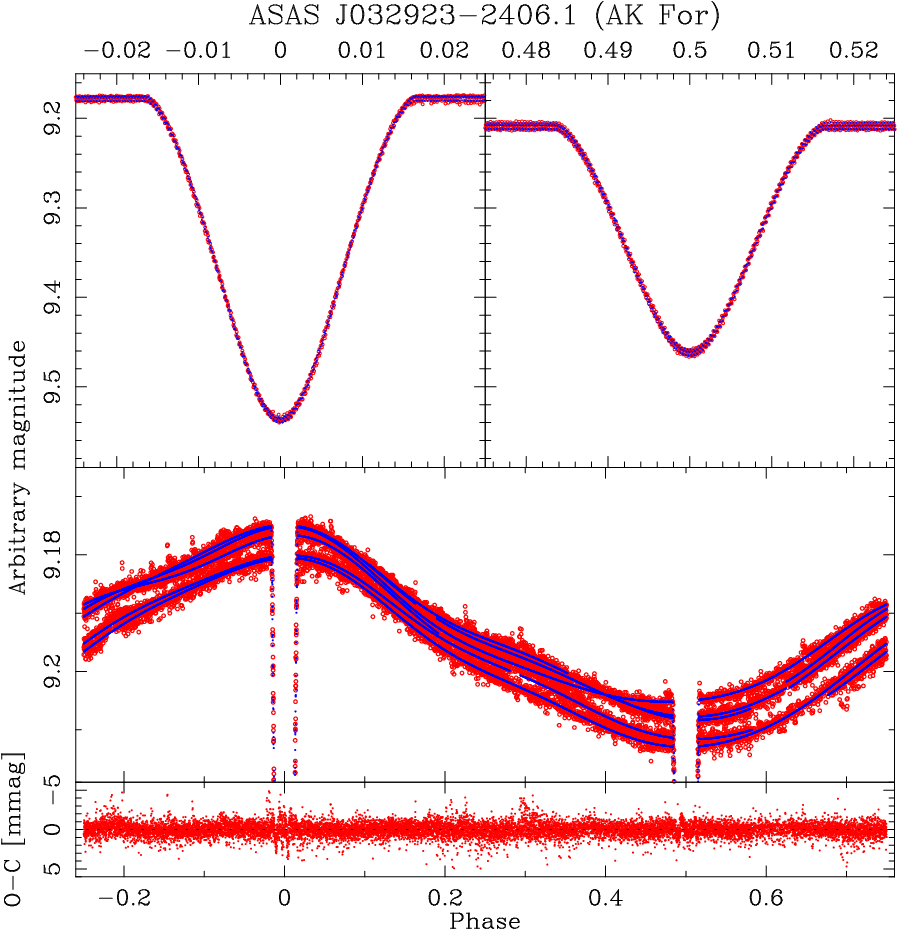}
\includegraphics[scale=0.43]{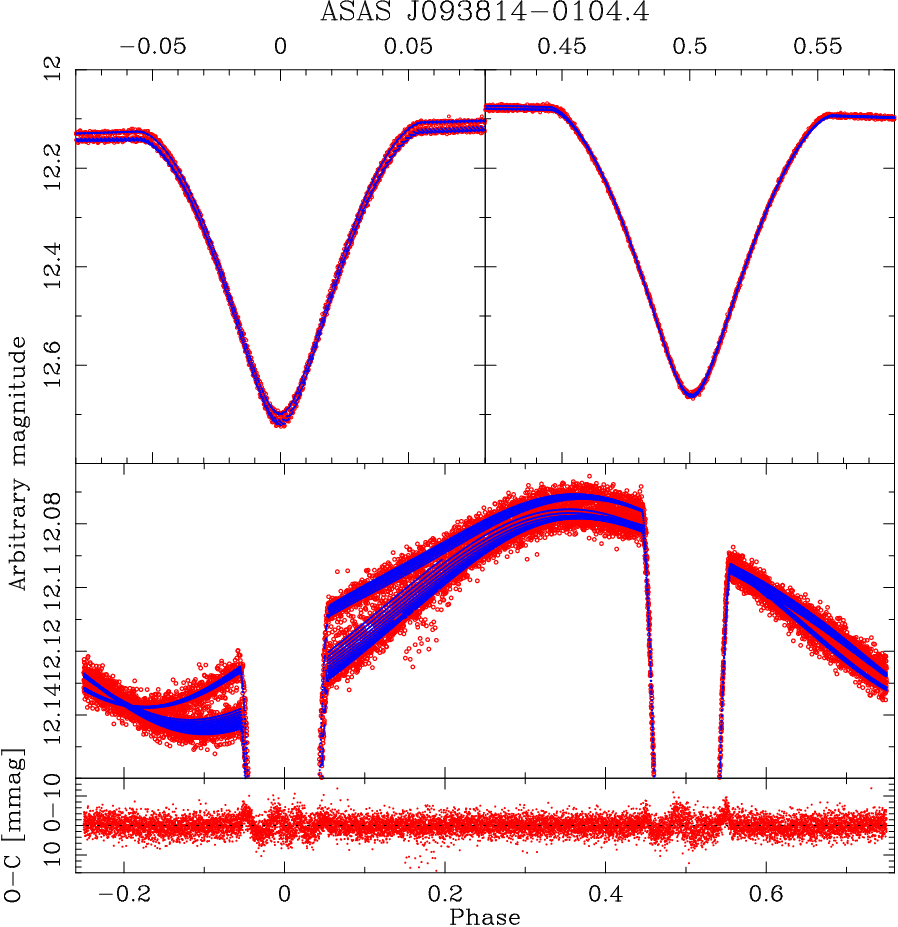}
\includegraphics[scale=0.43]{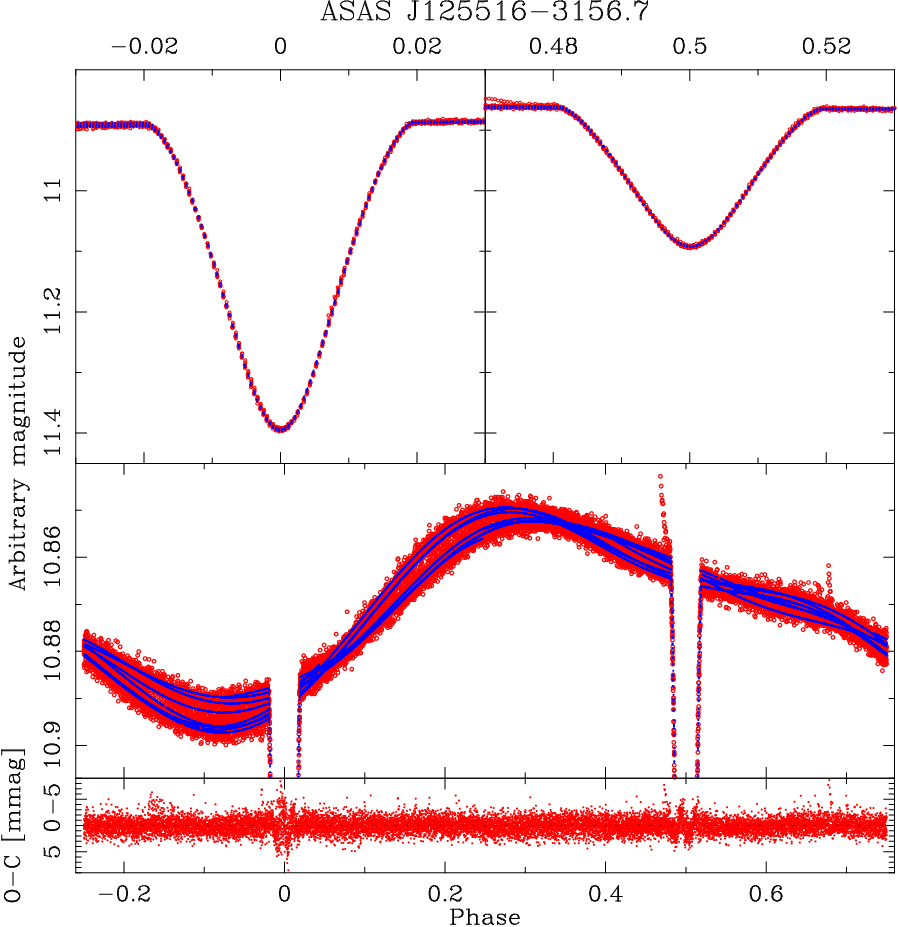}
\caption{TESS data (red) and JKTEBOP models (blue) of the studied systems,
phase-folded with orbital periods. Top rows are zooms on primary (left) and 
secondary (right) eclipses. Below are zooms on the out-of-eclipse modulations.
Bottom panels depict the residuals. One can clearly see the evolution of spots 
in time, as well as flares on AE~For and A-125.}\label{fig_lc}
\end{center}
\end{figure}

In order to obtain high accuracy and precision in stellar parameters, especially radii, 
one needs a very precise photometry, and the best-quality data come from space borne instruments.
The high-precision, 2-minute-cadence time-series photometry of our targets comes from the 
{\it Transiting Exoplanet Survey Satellite} (TESS), and was obtained through the Guest investigator program 
No.~G011083 (PI: He{\l}miniak) during the first year of TESS operations. Detrended light curves were 
downloaded from the Mikulski Archive for Space Telescopes (MAST). Our targets were mostly observed 
in one sector, except for A-011 (two sectors). TESS light curves are presented in Figure~\ref{fig_lc}.

\section{Spectroscopy and radial velocities}

Direct determination of masses of DEBs requires radial velocity (RV) measurements, which are obtained 
from a series of high-resolution spectra. Our targets were initially included into a large spectroscopic 
survey of DEBs, identified by the All-Sky Automated Survey (\cite[ASAS; Pojmański 2002]{poj02}). 

\begin{figure}
\begin{center}
\includegraphics[scale=0.43]{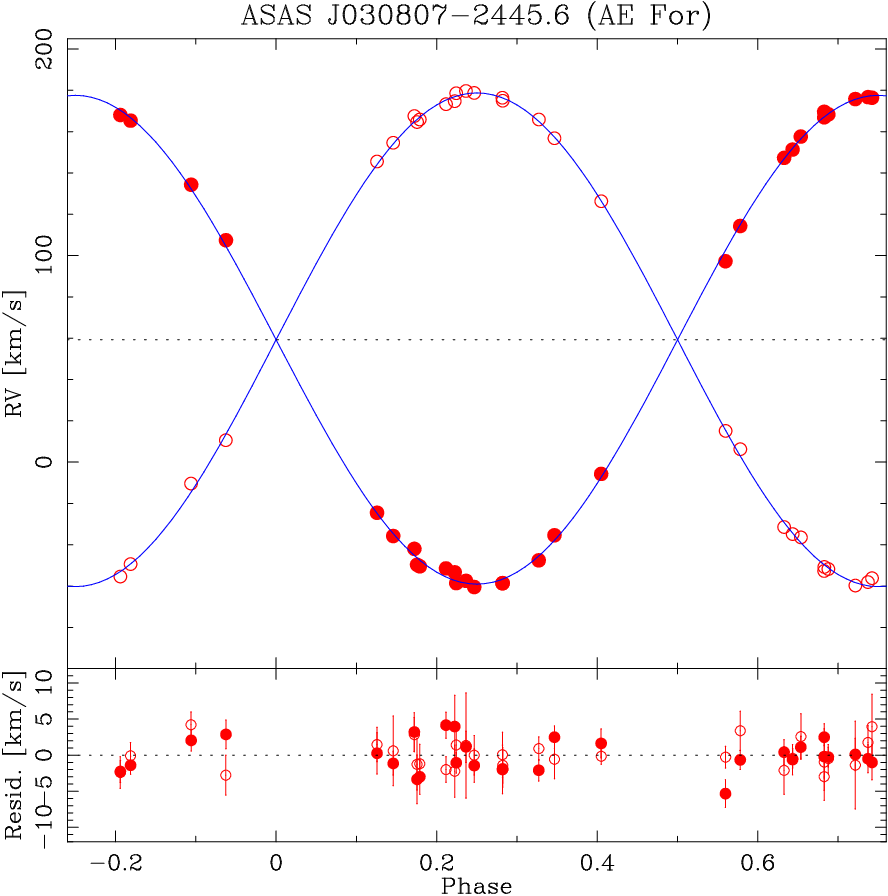}
\includegraphics[scale=0.43]{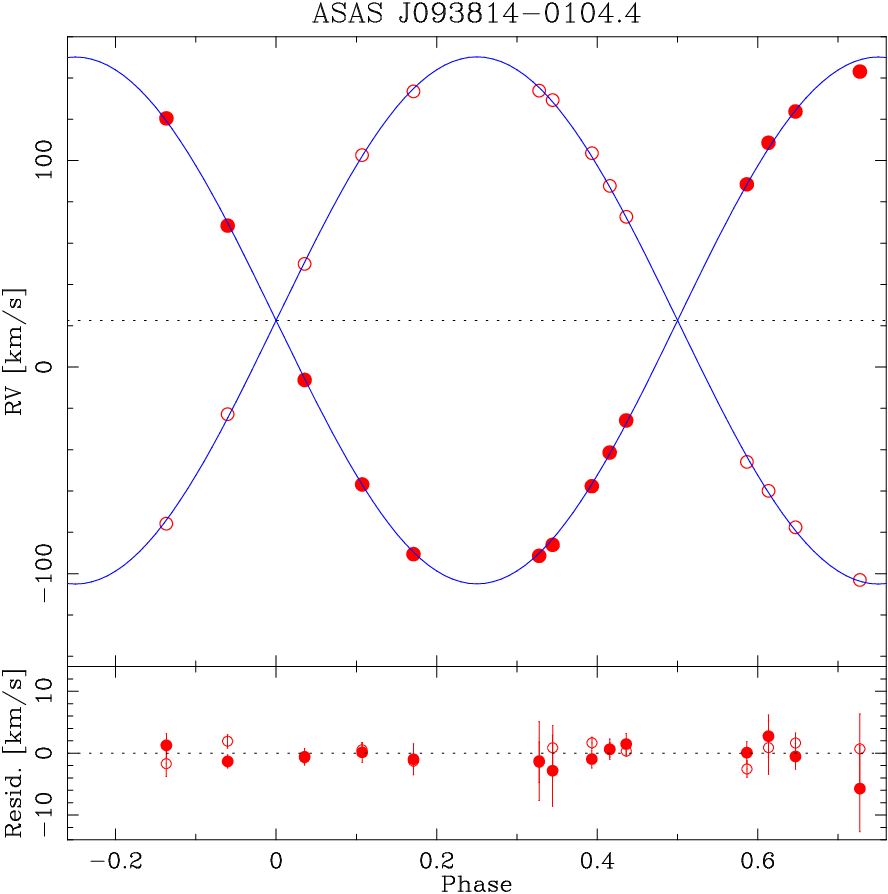}
\includegraphics[scale=0.43]{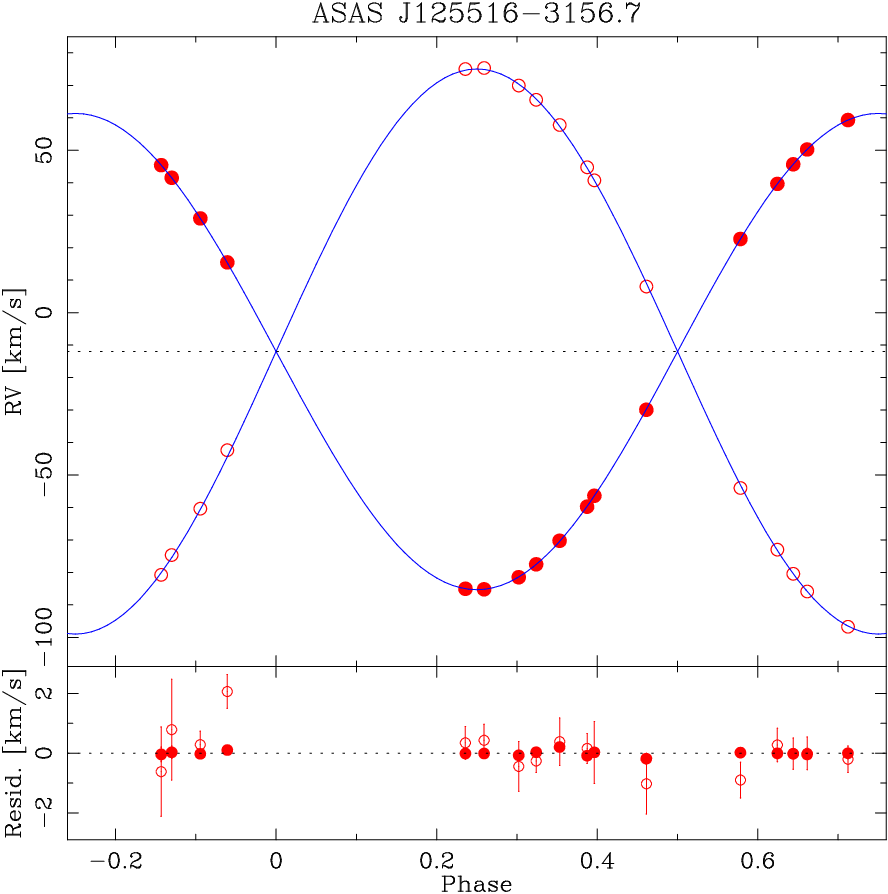}
\caption{RV measurements (red) and model curves (blue) of three systems with our CORALIE and CHIRON 
observations, phase-folded with their respective orbital periods. Corresponding residuals are shown 
in lower panels. Solid symbols represent data for primaries, while open for secondaries. Four
points with largest error bars in A-093 are data taken from \cite{hel11}.}
\label{fig_rv}
\end{center}
\end{figure}

RVs and orbital solutions of AK~For and A-011 remain unchanged with respect to the literature 
(\cite[He{\l}miniak \etal\ 2012, 2014]{hel12,hel14}). In three other cases we used our own new 
spectroscopy from CHIRON and CORALIE spectrographs, and calculated the RVs with the TODCOR 
method \cite{zuc94}. The CHIRON data for A-093
were supplemented with measurements from \cite{hel11}. 
AE~For was already described in \cite{roz13}, but we did not
use their data. The RVs for A-125 were not published to date. The orbital parameters were found 
with the code V2FIT (\cite[Konacki \etal\ 2010]{kon10}). The observed and model RV curves 
of AE~For, A-093, and A-125 are shown in Figure~\ref{fig_rv}.

\section{Light curve modelling}
The TESS light curves were modelled with the JKTEBOP code v34 (\cite[Southworth \etal\ 2004]{sou04}). 
To account for the out-of-eclipse modulation coming from spots we applied (in JKTEBOP) a series of 
sine functions (up to four) and polynomials (up to fifth degree). Because the spot-originated variation 
may change in time quite rapidly, data were split into several (between 2 and 6) pieces, which were 
analyzed separately. Parameter errors for each piece were evaluated with a Monte-Carlo procedure. 
As the final values we adopted weighted averages, and to get final parameter uncertainties, we added 
in quadrature a median of individual piece errors and the {\it rms} of individual results. The JKTEBOP
models are shown as blue lines in Figure~\ref{fig_lc}.

\section{Results}
In Table~\ref{tab} we present the most important results of our analysis, including RV semi-amplitudes
$K$, orbital period $P$ and inclination $i$, and absolute values of masses $M$ and radii $R$.

The variations in spot pattern, in time scales of single weeks, is the main difficulty in reaching 
good precision in radii. The behavior of residuals during the eclipses reflect the asymmetries 
and deviations of the shape of an eclipse from a ``clean photosphere'' case, and originate
from spots on a surface of the eclipsed component. However, thanks to the TESS data, we were able to
successfully model the influence of spots, and the uncertainties in radii are few times better than 
reported in literature. The exception is AK~For, where the ratio of radii 
$R_2/R_1$ is strongly correlated with the level of third light contamination. 

Spots also hamper the RV measurements, introducing additional jitter to the data. Also, components of
the shortest-period ($P < 1$~d) systems rotate rapidly. Nevertheless, new mass determination is also quite good.
Errors in masses for AE~For are quite low ($<$1\%), yet larger than those from \cite{roz13}, which 
is probably due to larger amount of their RV measurements. The new mass uncertainties of A-093 are 
2-3 times better than in \cite{hel11}, at the level of 1.5\%.
For the new system A-125 all properties are derived with high precision. 

Introduction of high-precision TESS photometry allows to improve our knowledge on the smallest, most
active stars, where magnetic fields and rotation strongly influence the observed properties. The five binary
systems presented here are only a sample of $\sim$40 DEBs with K- and M-type components observed by TESS
in our GI programs. Publications of the first set of final solutions is scheduled for mid-2020.

\vspace{0.3cm}
{\bf Acknowledgments:} K.G.H. acknowledges support provided by the Polish National Science Center through grant 2016/21/B/ST9/01613. N.E. would like to thank the Gruber Foundation for its generous support to this research.

\end{document}